\documentclass[12pt,dvips]{article}
\usepackage{epsfig}
\usepackage{hyperref}

\begin{document}

\title{Physical Origin of Elementary Particle Masses}
\author{Johan Hansson\footnote{\href{mailto:c.johan.hansson@ltu.se}{c.johan.hansson@ltu.se}} \\
 \textit{Division of Physics} \\ \textit{Lule\aa University of Technology}
 \\ \textit{SE-971 87 Lule\aa, Sweden}}

\date{}

\maketitle

\begin{abstract}
In contemporary particle physics, the masses of fundamental particles are incalculable constants,
being supplied by experimental values.
Inspired by observation of the empirical particle mass spectrum, and their corresponding physical interaction couplings, we propose that the masses of elementary
particles arise solely due to the self-interaction of the fields
 associated with the charges of a particle.
 A first application
 of this idea is seen to yield correct order of magnitude predictions for neutrinos, charged leptons and quarks.
We then discuss more ambitious models, where also different generations
may arise from \textit{e.g.} self-organizing bifurcations due to the underlying
non-linear dynamics, with the coupling strength acting as
``non-linearity" parameter. If the model is extended to include gauge bosons, the photon is automatically the only fundamental particle to remain massless as it has no charges. It results that gluons have an effective range $\sim 1$fm, physically explaining why QCD has finite reach.

\end{abstract}
\newpage
\section{Introduction}
The biggest, and oldest\footnote{I.I. Rabi's famous reaction to the discovery of the muon in the 1930s: ``Who ordered that?"}, unresolved enigma in fundamental particle physics is:
\textit{Where do the observed masses of elementary particles come
from?}, the concept of mass is not really understood, and their numerical values remain a mystery.

There is the widespread, but erroneous, belief that when the Higgs
boson is confirmed, the origin of mass has been found. This is
not the case. It merely replaces one set of unknown parameters
(particle masses) with an equally unknown set of parameters
(coupling constants to the Higgs field(s)), so nothing is gained
in the \textit{fundamental understanding} of masses.

Despite what can be imagined about the triviality regarding
masses from frequent statements like ``the supersymmetric
partners differ {\it only} in their masses (and spin) compared
to the normal particles", ``...the Planck scale,
$M_{Planck} = 1.22 \times 10^{19}$ GeV ",  and
``the mass of the top quark is
175 GeV", etc, the concept of mass has never been defined
in an unambiguous way, not even in classical physics.
In fact, Jammer
\cite{Jammer} has been able to write two whole monographs on this thorny
subject, concluding that nobody knows what mass really is.
Furthermore, and this is a problem even for the most
pragmatic of physicists, the mass parameters experimentally
measured for elementary particles have no theoretical
explanation whatsoever. From the vantage point of theory
the masses could just as well be a set of randomly
generated numbers.

To quote Richard Feynman: \\
``...although people say that there are no experiments to lead us,
it's not true. We have some twenty-four or more - I don't know the
exact number - mysterious numbers associated with masses. Why is
it that the mass of the muon compared to that with the electron is
exactly 206 or whatever it is, why are the masses of the various
particles such as quarks what they are? All these numbers, and
others analogous to that - which amount to some two dozen - have
\textit{no} explanations in these string theories - absolutely
none! There's not an idea at the present time, in any of the
theoretical structures that I have heard of, which will give a
clue as to why those masses are what they are...When you look at
these numbers, they look absolutely random and hectic; there
doesn't seem to be much pattern in them. That's a problem for
theoretical physics, and these superstring theories don't address
it at all." \cite{FeynmanString}
\\
 ``Throughout this entire story there remains one especially
unsatisfactory feature: the observed masses of the particles, $m$.
There is no theory that adequately explains these numbers. We use
the numbers in all our theories, but we don't understand them -
what they are, or where they come from. I believe that from a
fundamental point of view, this is a very interesting and serious
problem." \cite{Feynman}, p.152.

The fundamental concept of mass is still so poorly understood that it
was considered worthwhile to test whether an anti-atom
(produced at CERN for the first time  \cite{CERN})
falls up
or down in a gravitational field \cite{Harwit}.

Mass was ``smuggled" into the framework of quantum
mechanics, being the same parameter as in classical physics, {\it
i.e.}, the proportionality, or inertial, constant between the acceleration of a
body and the applied force. However, as neither force nor
 acceleration are appropriate
concepts in quantum mechanics, except through Ehrenfest's theorem
\cite{Ehrenfest}, as statistical or ``ensemble" entities, this
direct translation of the concept is somewhat dubious. In quantum mechanics and quantum field theory the concept of mass as ``resistance against acceleration" is simply inappropriate. The most
legitimate definition of mass in quantum mechanics seems to be by
the  relation $E = m$, or more correctly $m =$  $\langle E
\rangle$, {\it i.e.}, as something empirically
measurable, {\it e.g.}, by annihilation/production of
particle-antiparticle pairs.

As all other {\it observables}  in quantum mechanics,
mass should
be represented by an {\it operator} with its resulting
smeared out probability distribution (unless an exact eigenstate), peaked at the classical
value.
Naively, the only way to get
discrete values for the mass would be to assume that the
``elementary" particles are bound states of something more
fundamental \cite{preons}, \cite{PreonTrinity}. Probably the
only way to end this
infinite regress to smaller scales, would be to assume that the
ultimate
subconstituents are massless. The mass would then necessarily
have to arise strictly from the dynamics, ``mass without mass", John Wheeler's goal to remove mass from the \textit{basic} equations of physics \cite{Wheeler} - as implicitly anticipated by Einstein in his article \textit{Does the inertia of a body depend upon its energy content?}: ``The mass of a body is a measure of its energy-content" \cite{Einstein}.

According to ``Mach's principle" \cite{Mach} the mass
of a particle arises
because of its interaction with the rest of the universe.
This would mean that mass would no longer be a scalar
but a tensor (of second rank), as inhomogeneities and
anisotropies in the distribution of surrounding matter would
give a directional and temporal dependence in the inertial mass. This is
accidentally
similar to the modern parametrization of mass as a matrix (tensor of
second rank) when discussing fermion masses \cite{Fritzsch}
in general, and neutrino oscillations in particular.

As none of the speculative theoretical ``advances'' since the 1970s, such as
grand unified theories (GUTs), supersymmetry (SUSY), superstrings
and M-theory to name but a few, has yet yielded a single
experimentally tested result, we instead ask if the mass spectrum can be
understood (and derived) from what is known to be more or less
correct, {\it i.e.} the Standard Model of particle physics.
Although some proponents of more ``fundamental'' theories claim to, eventually, be
able to deduce everything from first principles, including the mass spectrum of
elementary particles, there has not yet been a single instant
where this has been achieved in practice, and we thus prefer a more
``pragmatic'' approach that can be initiated {\it
today} instead of waiting for the (perhaps nonexistent)
``ultimate'' theory to tell us all truths.

Can we understand the different particle masses from
what we already know, more or less, to be correct, {\it i.e.} without
any ``exotic'' physics? We believe that this may be possible,
and will elaborate on that in what follows.

We propose that the mass of a particle has a strictly
local origin (not a global one like \textit{e.g.} in the assumption of Mach), arising from its self-interaction(s). That is, the mass is equivalent to the energy contained in the associated gauge fields (in perturbative quantum field theory; the energy of the ``cloud" of virtual gauge particles). Such a connection between fundamental dynamical interactions and mass seems only reasonable as only mass is needed to go from kinematics to dynamics \cite{Jammer}. As most of the (at least) eighteen arbitrary parameters of the ``standard model" arise because of the mass problem, a connection between the masses and the ordinary, non-Yukawa \textit{i.e.} non-Higgs, interaction couplings would also significantly reduce the number of free \textit{ad hoc} parameters.

\section{Perturbative model}

In the conventional approach to particle masses within the
standard model (and beyond), the Higgs field generates the
different masses through its coupling strength to the other
fields. These couplings are arbitrary parameters, chosen to
coincide with experimental data. So even if the Higgs model turns
out to be the correct one to break the electroweak symmetry (which
theoretically seems somewhat improbable as the original Higgs then would be the only
spinless fundamental field, with all the theoretical problems that
entails) nothing is gained in the knowledge of why the elementary
particles have the masses we observe. The Higgs mechanism simply
replaces one \textit{ad hoc} mass parameter ($m_i$) with another
equally \textit{ad hoc} Yukawa coupling constant ($\lambda_i$) to the Higgs field
\begin{equation}
m_i {\leftrightarrow} \lambda_i,
\end{equation}

and because of this we are free to make other hypotheses as to what \textit{physically}
generates mass. A Higgs-like symmetry breaking mechanism is the favorite way to break also other (hypothetical) symmetries,
such as supersymmetry. To us that seems to be the wrong way to
proceed, as such a symmetry-breaking mechanism always introduces
new free (\textit{ad hoc}) parameters, and a more fundamental theory should
contain \textit{fewer} free parameters, not more. As previously stated, the original standard
model itself contains 18 free parameters, and several more if
neutrinos are non-massless, most of them related to the
theoretically incalculable masses. In supersymmetric extensions of
the standard model the free parameters rank in the number of
hundreds, or more, again mainly connected to new (unobserved) masses of
supersymmetric partners, and to new Higgs-like mechanisms at
higher (incalculable) energy scales.

Here we, instead, start by assuming that {\it all} of a particle's
mass arises exclusively from its interaction with itself (see
Fig.1). This makes it possible for the underlying ``bare''
lagrangian to include only massless fields, just like in the Higgs model before the symmetry is broken, preserving the
gauge-invariance of the standard model, as the self-coupling is a
vacuum phenomenon, just like the non-vanishing vacuum expectation
value of the Higgs field when the symmetry is broken at ``low" energies. Mass is thus a ``frozen" irreducible energy connected to, and defining, the particle.
\begin{figure}
\begin{center}
\includegraphics {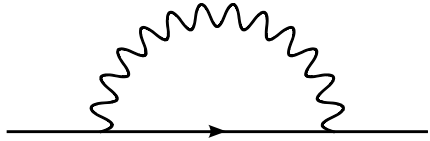}
\end{center}

\caption{The $O(\alpha)$
quantum field theory contribution to the mass
of a particle due to self-interaction. Each vertex contributes one charge factor $\sqrt{\alpha} \propto q$, so $m \propto \alpha$; the mass is proportional to the \textit{physical} coupling constant.}
\label{Fig1}
\end{figure}
In a \textit{perturbative} quantum field-theoretic sense, the diagrams for self-interaction
are of course divergent due to virtual
loops, but surely nature herself is not singular. It seems obvious that the formally infinite masses predicted by perturbative quantum field theory is only an artifact of the perturbative approximation. (Even the conventional ``solution", the renormalization of masses and charges, is defined through the approximate, perturbative formulation of the theory.) Furthermore, as all gauge fields, \textit{i.e.} different interactions, in the standard model diverge in the same way \textit{quotients} between masses are finite. There seems to be little reason to
work at some very high (GUT/String) scale, as the masses
we are
interested in
are observed at ``normal'' energies. This also
spares us of evolving the behaviour at extreme ``fundamental" energies
down to experimentally accessible energies via the
renormalization group.
The renormalized, {\em i.e.}, physically
measurable, mass arising
from the altered propagator due to higher
order perturbative corrections in quantum field
theory can be written
\begin{equation}
m = m_0 \, ( 1 + \sum_{n = 1}^{\infty} \alpha^n d_n (\Lambda^2) )
= m_0 + \delta m,
\end{equation}
where $m_0 $ is the ``bare'' (non-observable)
mass, and $d_n (\Lambda^2)$  ``ultraviolet" diverges
as the spatial cut-off $\Lambda \rightarrow 0$.
Our approach is somewhat like taking the
limit $m_0  \rightarrow 0$ and $\Lambda
\rightarrow 0$ simultaneously.

The first obvious conclusion is that
the stronger the self-interaction,
the more massive
the particle will be.
The neutrino, interacting only through the
weak force, will become very light. Next,
the electron with electromagnetic coupling,
becomes more massive. The quarks, interacting also through the strong force, become heavy.

To illustrate the point more clearly,
we construct
a simple model.
The detailed non-perturbative dynamics will most certainly alter the
(perturbatively formally infinite) proportionality constant for the
different interactions (due to different gauge groups),
which in the ratio of particle masses
hence will not identically
cancel, but give a remaining finite factor.
For now, to obtain first approximations,
we disregard this complication and assume it to be of order unity.
For the mass of a particle (to order
$O(\alpha)$)
we thus make the ansatz

\begin{equation}
m = B \, (Q^2  \alpha_{em} +
T^2  \alpha_{weak}
+ C^2  \alpha_{s} ),
\end{equation}
where $\alpha_{em}$ is the
Sommerfeld fine structure
constant ($\sim 137^{-1}$)
and $\alpha_{weak}$
and $\alpha_s$ are the (low-energy)
couplings of the weak-,
and strong interactions.
As $\alpha_{weak}$ is ambiguous
(mass dependent), we infer the effective $\alpha$'s
from the measured
characteristic reaction times
({\em e.g.}, decay probabilities), as the interaction
strength is but a measure of the probability
of the interaction to take place.
 $Q$ gives
the particle's electrical charge in units of $e$,
and $T$ and $C$ are analogous quantities (of order one) coming from the gauge-groups for
``weak charge'' and ``color charge". $B$ is a normalizing
constant, which in a truly non-perturbative treatment of the standard model should be calculable. However, using quotients, it cancels.

This gives the results
\begin{equation}
m_{\nu}  /
m_e \sim  10^{-7} ,
\end{equation}
\begin{equation}
m_q  / m_e \sim  10^2,
\end{equation}
or in mass
units
\begin{equation}
m_{\nu} \sim 0.1 \, eV,
\end{equation}
clearly compatible with direct experiments; $m_{\nu_e} < 2 $ eV \cite{Part Prop}. Right-handed (sterile)
neutrinos, if they exist, would
be strictly massless in this scheme as they have no self-interaction.
\begin{equation}
m_q \sim 10 \, MeV,
\end{equation}
compatible with current quark masses \cite{Part Prop}.

If couplings of higher order
are taken into
account, the electromagnetic and weak
contributions will essentially be unchanged,
as they give rapidly diminishing terms in the perturbative expansion.
The higher orders in $\alpha_s$, however,
give a perturbatively divergent result, and hence an ``infinite"
quark mass, in accordance with some
``explanations'' of quark confinement
\cite{GellMann,Close}.
If quarks are permanently confined, it is troublesome
to even define a mass for them, as Wigner's mass-spin
classification of elementary particles \cite{Wigner} (unitary representations
of the Poincar\'{e} group) is valid strictly only for
free particles (asymptotic states).

\section{``Improved'' perturbative model}
R.P. Feynman: ``This repetition of particles with the same properties but
heavier masses is a complete mystery." \cite{Feynman}, p.145.

The three different particle ``generations''
of the standard model can be
accounted for (if not explained)
by a straightforward
generalization
of the formalism,

\begin{equation}
m = B_i \, (Q^2 \alpha_{em} +
T^2  \alpha_{weak}
+ C^2  \alpha_{s} ),
\end{equation}

where $B_i$ denotes the normalizing constants
for the three different generations, $i \in 1, 2, 3$.

If one introduces a ``generation charge'' or quantum number, $G_i$, the $B_i$
could be written $B_i = B \, f(G_i)$, where $f(G_i)$ is
a function of $G_i$.

This gives the results (taking the readily measured $m_{\mu}$
and $m_{\tau}$ as known):

$m_{\nu_{\mu}} \sim 10$ eV,
$m_{\nu_{\tau}} \sim 100$ eV, well
below, and hence consistent with, the direct experimental upper limits of
$170$ keV and $15.5$ MeV
\cite{Part Prop}.

A physical mechanism for
connecting the different $B_i$'s
would be highly desirable, see section 5 below.
Still, the number of arbitrary parameters has decreased compared to the orthodox,
``masses-from-Higgs''-mechanism \cite{Cargese} due to the
relation between coupling strengths and masses,
which in the
standard model are completely independent quantities.

If we enlarge the model to include a tentative fourth generation, and use the most simple power-law ansatz, $f(G_i) = G_i^{\beta}$, utilizing the known values for the three charged leptons corresponding to $G_i \in 1,2,3$ to make a curve-fit resulting in $\beta \simeq 7.4$, we get a prediction for the mass of an additional charged lepton $\kappa^-$ \cite{PreonTrinity}, with $G_i = 4$, as $m_{\kappa} \simeq 20$ GeV, which seems ruled out as the experimental limit for a heavy charged lepton is $> 100$ GeV \cite{Part Prop}. So a more elaborate form for $f(G_i)$ should be sought. However, as this would provide no \textit{real} fundamental understanding, merely a parametrization, we refrain from pursuing this avenue further in this article.

\section{Dimensional analysis}
An even simpler picture can be deduced by pure dimensional analysis.
If we assume that mass is directly, and solely, related to the couplings,
and that the standard model is essentially ``correct'', we
only have three couplings to utilize.
These are the fine structure constant, $\alpha_{em}$, the strong
interaction ``constant'',
$\alpha_s$, and the Fermi constant, $G_F$.
To get a mass scale, we have no choice but to use $G_F^{-1/2} \sim 300$
GeV, as this
has the dimension of mass (when $c = \hbar = 1$) whereas the other two are dimensionless.
The formula for mass then takes the form

\begin{equation}
m = G_F^{-1/2} f(\alpha_{em}, \alpha_s),
\end{equation}
where $f$ is a dimensionless function of only the fine structure
constant and the strong coupling.

As the top quark mass $m_t$ is of the same order of magnitude as $G_F^{-1/2}$ this could mean that no more massive quarks exist and, if the general structure is a true facet of nature and not just an artifact of the standard model, only three particle generations exist.

\section{Nonperturbative models}
Seeing that even our first-order
model
for the masses of elementary particles
gives results of the right order of magnitude,
we are now ready to discuss something more
sophisticated.



Truly nonlinear effects have so far received little attention in particle physics, as the dominating and inherently \textit{perturbative} Feynman diagram techniques mask the nonlinearities and tend to give behavior that closely mimics the linear case (although interactions are perturbatively, \textit{i.e.} mildly nonlinear). We believe that the utilization of nonlinear methods, as developed in other fields of science, could greatly benefit elementary particle physics, especially in the case of masses. The most typical feature of nonlinear equations is multiplicity of solutions, raising a hope that different generations may be automatic. In the contemporary understanding of the standard model these repetitions of particles is a complete mystery.

Interacting quantum field theories are really inherently non-linear, as can
easily be seen by an analogy with classical waves. Two waves
meeting in a perfectly linear medium simply
penetrates each other
unaffected (think two laser beams in vacuum, or $\sim$ air). This is the equivalent of non-interacting
quantum field theory. A necessary
condition for the waves to interact
({\em e.g.}, scatter) is that the medium is non-linear.
We could thus use a classical analogy to guide our thinking
in constructing a phenomenological model.
This would mean to view the vacuum as a non-linear reactive
medium. Particles with different charges
would thus ``see" a different (effective)
medium and react differently, generating different masses.

In more mathematical terms, this can within the standard model be described by the two
coupled evolution equations for gauge ``force fields", $F_{\mu \nu}$
and ``matter fields'', $\psi$.

\begin{equation}
D_{\mu} F^{\mu \nu} = g \bar{\psi} \gamma^{\nu} \psi,
\end{equation}

\begin{equation}
(i D_{\mu} \gamma^{\mu} - m) \psi  = 0,
\end{equation}
where
\begin{equation}
D_{\mu} = \partial_{\mu} - i g A_{\mu},
\end{equation}
is the covariant derivative, $g$ the coupling constant (the ``nonlinearity parameter", essentially $\sqrt{\alpha}$) and $A_{\mu}$ the gauge field
potentials. The gauge field strength tensor is given by
\begin{equation}
F_{\mu \nu} = \frac{i}{g} [D_{\mu}, D_{\nu}] = \partial_{\mu} A_{\nu}
- \partial_{\nu} A_{\mu} -i g [A_{\mu}, A_{\nu}] .
\end{equation}

We see that, even for an abelian theory like QED, where the commutators $[A_{\mu}, A_{\nu}]$ vanish (physically reflecting that the photon has no self-interaction),
the system of equations (10), (11), constitute a nonlinearly
coupled system. In fact, in Eq.(11) we would like to take $m=0$
(or rather $m=\epsilon$ infinitesimal due to quantum fluctuations)
from the outset and {\it deduce} a scalar factor of this form
({\it i.e.} a mass) from the coupled system.

Traditionally, interactions in quantum field theory are treated
as small perturbations around the non-interacting background
state (Feynman diagram method, etc).
This is evidently applicable only to mildly non-linear theories,
{\em e.g.}, theories in which the coupling constant, $g$,
 is small, as for
instance in quantum electrodynamics. For a theory of masses,
however, the exact formulation is required, as a truncated
perturbative series does not capture the kind of non-linearity
we are interested in. Furthermore, the very act of (perturbative)
renormalization
destroys all information about any possible underlying
mechanism for mass generation, as the real (measured) masses
must be taken as experimental input. So, it seems obvious that
a perturbative \textit{fundamental} understanding of masses is impossible. One has
to take into account that nature seems to work wholesale
and does ``everything at once'',
not piecemeal as implied by Feynman diagrams ({\it i.e.} by a
perturbative expansion of an otherwise hitherto intractable problem).
\subsection{Solitons}
It is well known that nonlinear theories may have particle-like
solutions. There are many known ``solitary wave solutions", where
the normal dispersive effect is exactly counterbalanced by a
nonlinear focusing effect, giving waveforms that are unaltered. If
the solitary waves also are unaffected by collisions with other
solitary waves they are called ``solitons". Soliton solutions thus in many ways
act like particles. There are many known examples of analytical soliton
solutions for 1+1 dimensional systems (Korteweg-de Vries, sine-Gordon, etc), much fewer for 2+1, and none for relativistic 3+1 dimensional systems with non-trivial soliton scattering. As quantized solitons arising in relativistic field theories possess many of the attributes of particles, such as mass, charge and spin \cite{Manton}, it would
be very tempting to identify elementary particles with exact
solitary wave solutions to the 3+1 dimensional quantum field
theory of the standard model, or its (unknown) dual \cite{Manton}, and stable elementary particles with
soliton solutions to the same equations. So far there is no integrable model with solitons respecting the symmetries in Minkowski space-time. (An example of an integrable quantum field theory, believed to bear a close similarity to the four-dimensional Yang-Mills theory of the standard model, is the two-dimensional nonlinear sigma model.) Numerical studies of non-integrable models show that solitons do scatter, soliton-antisoliton collisions lead to annihilation, the energy dissipating into wave-like solutions of the linearized field equations, sometime after transient periods in which the energy localizes into wavepackets called ``oscillons". This makes it tempting to identify soliton-antisoliton annihilation as \textit{particle-antiparticle} annihilation, oscillons with secondary (semi-stable) daughter/cascade particles, and the radiated energy as the final liberated energy from annihilation.

In that vein, Skyrme's old pioneering unified model of nucleons and nuclei \cite{Skyrme}, in terms of soliton ``Skyrmions", with almost no free parameters, seems a compelling inspiration (and goal) also for truly elementary particles.
\subsection{Self-Organization}
Self-interacting non-linear systems have been
studied for a long time in the theory of dynamical systems, chaos  \cite{Cvitanovic,BaiLin} and spontaneously self-organizing phenomena \cite{Prigogine}.
They are ideal for investigating the behaviour
of systems which self-interact over and over again,
hence making them useful models for our ideas about
mass-generation.
Moreover, there is a limiting
behaviour which gives
the same results for large classes of models,
regardless of dynamical
details (``universality") \cite{Feigenbaum}.
This limiting behaviour is not only qualitative, but
 {\it{quantitative}}. It is regarded as a fundamental
organizing principle of nature on the macroscopic scale,
but could in principle also apply to the quantum world
as elementary particle interactions are inherently non-linear as noted above.
Furthermore, the production and decay of unstable (heavy) elementary particles is physically a clear sign of far-from-equilibrium behavior, \textit{e.g.} $\mu \rightarrow e \bar{\nu}_e \nu_{\mu}$. Non-equilibrium and nonlinearity generally being regarded prerequisites for self-organization \cite{Prigogine}. As the vacuum is displaced further and further from equilibrium (\textit{i.e.} becomes ``excited"), more and more states become possible, see Fig 2.

We like to picture the elementary particles as
semi-localized ``knots" in the
fields permeating space,
arising because of the non-linear response of the field
due to its charge(s). Mass would then be just another example of an emergent phenomenon so characteristic of many nonlinear systems. For example, in solid-state physics, ``intrinsic localized modes" \cite{Dolgov}, \cite{Sievers} were surprisingly discovered in the late 1980s and are now known to be typical excitations in strongly nonlinear systems. As quantum field theory is also a nonlinear system, it does not take much imagination to expect analogous phenomena to occur also in particle physics. So instead of particle masses arising from \textit{extrinsic} sources (brute-force, linear, \textit{ad hoc} Higgs couplings) they could be due to \textit{intrinsic} sources (automatic, nonlinear, physical self-interactions).


It is perfectly possible to get phenomena of great complexity even with a remarkably simple underlying setup. Particle masses may be just yet an example of this very general phenomenon.
The self-interaction ``mapping" cannot be linear, as this would lead to infinite
(exponentially increasing) or vanishing (exponentially decreasing)
masses. One of the simplest nonlinear equations fulfilling these criteria is the celebrated ``logistic mapping''

\begin{equation}
x_{t+1} =  k x_t (1 - x_t) ,
\end{equation}
which still leads to surprisingly complex dynamics \cite{May} and universality \cite{Feigenbaum}.


In order to indicate how masses could be constructed in such a setting, we use the logistic mapping as a ``toy model". The true dynamics, however, is
believed to be much more complicated, but the {\it qualitative}
features of the true mechanism could be similar, as even models governed by partial differential equations (formally $\infty$-dimensional dynamical systems; equivalent to dynamical systems of infinitely many coupled equations) can, very surprisingly, be well-described by low- or even one-dimensional nonlinear dynamical systems \cite{Cvitanovic}. ``Dissipation bleeds a complex system of many conflicting motions, eventually bringing the behavior of many dimensions down to one" \cite{Collet}.

We take $x$ to be proportional to the mass generated by the self-interaction of the field with itself ($x_t \rightarrow x_{t+1}$).

If we again use the Fermi-constant (the only dimensionful coupling constant in the standard model)
\begin{equation}
x = \frac{m}{G_F^{-1/2}},
\end{equation}
is a dimensionless mass parameter. The discreteness in time could be
justified by the possible existence of a smallest time-interval (\textit{e.g.} the
``Planck time'') but also from a pragmatic point of view as it makes calculations simpler.

The non-linearity parameter is a function of the dimensionless couplings, $k = k(\alpha_{em}, \alpha_s)$ and we start with $m_0 = \epsilon$ due to quantum fluctuation. The physical justification of Eqs. (14), (15) comes from Fig. 1. Each loop contributes $kx$, but at the same time nature can ``borrow" only a finite energy from the vacuum (the ``carrying capacity" in the logistic equation) taken as $G_F^{-1/2}$ in this case. So, physically, the first (loop) term in Eq. (14) leads towards the normal diverging self-mass contribution, whereas the second term tends towards ``starvation"/suppression. Heisenberg's uncertainty principle, due to the finiteness of $\hbar$ and the discrete time-step, means only a finite energy is available from the vacuum.

The logistic mapping, for small $k$, has the behavior:

$k < 1 \Rightarrow
m \rightarrow 0$. This would then be interpreted that too weak self-interactions will result in strictly massless particles (\textit{e.g.} gravitons).

For $1 < k < 3 \Rightarrow m_{t+1} = m_t =
m^*$, \textit{i.e.} point \textit{attractors}, giving single asymptotically stable solutions for intermediate couplings.

$3 < k < 3.57... $ give multiple solutions, perhaps identifiable as generations. As the coupling $k$ increases, previously stable solutions become unstable (repellors) while the new bifurcation solutions become new stable states (attractors). Stable particles could thus correspond to fixed points ($m_{t+1}
= m_t$) which are {\it attractors} while unstable
particles may correspond to fixed points which
are {\it repellors}.

$3.57... <
k < 4 \Rightarrow$ chaotic solutions, interspersed with infinitely many narrow windows of multiple solutions. This might be connected to the suggestion that nature is chaotic at the basic fundamental level, but goes through self-ordering at higher (observable) levels \cite{Froggatt}.

\begin{figure}
\begin{center}
\includegraphics {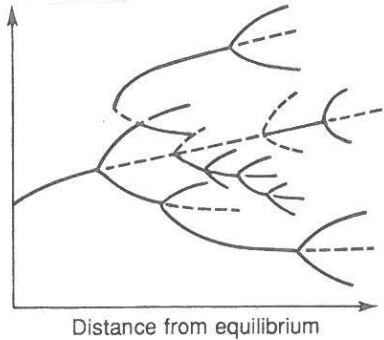}
\end{center}

\caption{Schematic plot of generic bifurcation diagram, in our case Mass vs Distance from equilibrium, arising from spontaneous self-organization - with many complicated ``forks" as we go towards the right. As the distance from equilibrium increases, a multiplicity of stable states typically become possible. Dashed lines indicate unstable states.}
\label{Fig2}
\end{figure}


When the coupling decreases, we see that we get
``unification", not necessarily of forces , but of
{\it masses}, implying that there exists only {\it one}
charged and massless lepton as $k(\alpha_{em}, \alpha_s)  \rightarrow 0$.
Furthermore, if the dynamics is governed by a
GUT/String-like theory, with only one primordial coupling
parameter, quarks and leptons are {\it the same}
(apart from hitherto unexplained quantum numbers).

As $0 \leq x \leq 1$ for the logistic equation, we see that elementary
particles, in this simple model, cannot have arbitrarily large masses, but that $m \leq 300$ GeV.
This would mean that the top quark ($x \sim 1$) is presumably the heaviest
elementary particle that exists.

\section{Gauge bosons}
If we assume the model in section 2 to also apply for gauge bosons, we see that only electromagnetism (quantum electrodynamics) has truly infinite reach; as the photon carries no charge it has no self-interaction and its physical mass remains zero.

The strong force (quantum chromodynamics) should disappear exponentially at sufficient distances due to the non-zero effective physical mass of its force carrier particles, the gluons, due to their self-interactions. The range can be estimated by the non-massless theory potential $e^{-\lambda m c/\hbar} /r$, giving $\lambda_{cutoff} \simeq \hbar/mc$. This gives for the gluon with bare mass zero (preserving gauge-invariance in the lagrangian), but physical mass $m(gluon) \neq 0$, the value $\lambda_{cutoff}(QCD) \simeq$ 1 fm. This explains why QCD is only active within nuclei, although the bare gluon mass $m=0$ naively would give infinite reach within the standard model as its coupling to the Higgs is zero - despite what many think, this problem has never been solved \cite{Clay}.

The very massive gauge bosons $W^{\pm}$
and $Z^0$
evidently pose a problem. In our scheme
$m_W \sim m_e$
and $m_Z \sim m_{\nu}$,
which of course
is ruled out. There are other theoretical
expectations
however, which point at the possibility
of $W$ and $Z$
being composite particles, in which
case their intricate
inner structure would set the mass scale
\cite{preons}, \cite{PreonTrinity}.

If we extrapolate outside the standard model to also include gravity, the
mass of the graviton, the hypothetical
carrier of at least
weak gravity \cite{boulware}, should be scale
dependent. The mass of a ``soft" graviton
would still be zero for all practical purposes,
while the mass of a highly energetic graviton
should be proportional to its frequency due to
self-interaction with its own field (as the ``gravitational charge'', \textit{i.e} mass,
increases with increasing energy).
Gravity would then be described by a
gauge theory similar to the weak interaction
(with massive spin-2 quanta instead of the
spin-1 quanta of the weak theory). As both the
Fermi coupling constant $G_F$ and Newton's
gravitational constant $G$ have dimensions
$(mass)^{-2}$ in units where $\hbar = c = 1$,
it is not inconceivable that the
incorporation of massive
intermediate field quanta, where the mass is \textit{not} put into the lagrangian ``by hand", could
render perturbative
quantum gravity tractable,
just as in the case
of the weak interaction.
The difference being that the mass of
the graviton would be scale-dependent,
whereas the $W$ and $Z$ in the standard model are taken to have
constant masses.
(Actually, {\em all} quanta would be affected
in a similar way, as
they all carry ``gravitational charge'', {\em i.e.},
energy-momentum, possibly giving an effective
cut-off to loop integrals, rendering all quantum field theories
ultraviolet finite.)
This would alter, in a dramatic way,
the behaviour of gravity at short distances,
giving an asymptotic effective coupling which
tends to zero for very high energies,
softening the ultraviolet divergences.
It would also
give the right limiting behaviour of classical
(very many
soft, low-energy gravitons) gravity, as only the
``soft'', essentially massless, gravitons can
escape an appreciable distance to
contribute to the
long range gravitational field.
The effective (non-relativistic)
gravitational potential would
then be modified to something like
\begin{equation}
V \sim \frac{e^{-m(\omega) c r/ \hbar}}{r},
\end{equation}
where $m(\omega)$ is the frequency dependent
effective mass of the graviton.
It would possibly also modify the {\em very} long range
classical behaviour of gravity,
making it essentially
non-infinite in reach, altering the long-range $1/r$-dependence, in accordance with some modifications
proposed to explain galactic dynamics without
``Dark Matter'' and cosmological expansion without
``Dark Energy''.

\section{Conclusions}
We have seen that it might be possible to understand, both qualitatively and quantitatively,
much of the fundamental particle spectrum in a pragmatic way,
without turning to untested ``exotic'' physics. We have not proven anything, but then again physics is not about proofs but about correspondence with empirical facts. Still our scenario seems so simple, natural and physically compelling that we believe it to probably contain at least part of the real mechanism.

However, more refined and sophisticated models
of this kind
are needed before
anything conclusive can be said about this
proposition, although we feel
it is a step in the right direction to be
able to actually {\it derive} masses from
fundamental principles.
As essentially
nothing is known today about why elementary
particle masses have the values observed, each small step
towards a resolution
of this puzzle is worthwhile. The very first small steps towards the ideas presented in this article were described in \cite{Avhandling}.

\end{document}